\documentclass[preprint]{emulateapj}

\shorttitle{Spectro-photometric Search for Galaxy Clusters}
\shortauthors{Yoon et al.}
\slugcomment{Accepted by ApJS: December 4, 2007}

\begin{document}

\title{A Spectro-photometric Search for Galaxy Clusters in SDSS}

\author{Joo H. Yoon\altaffilmark{1,2},
Kevin Schawinski\altaffilmark{3},
Yun-Kyeong Sheen\altaffilmark{1,2},
Chang H. Ree\altaffilmark{2},
and Sukyoung K. Yi\altaffilmark{1,2}}

\altaffiltext{1}{Department of Astronomy, Yonsei University, Seoul 120-749, Korea}
\altaffiltext{2}{Center for Space Astrophysics, Yonsei University, Seoul 120-749, Korea}
\altaffiltext{3}{Department of physics, University of Oxford, Oxford OX13RH, UK}
\email{yi@yonsei.ac.kr}

\begin{abstract}
Recent large-scale galaxy spectroscopic surveys, such as the Sloan Digital
Sky Survey (SDSS), enable us to execute a systematic, relatively-unbiased search
for galaxy clusters. Such surveys make it possible to measure the 3-d
distribution of galaxies but are hampered by the incompleteness
problem due to fiber collisions.
In this study we aim to develop a density measuring technique that
alleviates the problem and derives densities more accurately by
adding additional cluster member galaxies that follow optical
color-magnitude relations for the given redshift.
The new density measured with both spectroscopic and photometric data
shows a good agreement with apparent information on cluster images
and is supported by follow-up observations. By adopting this new method,
a total of 924 $robust$ galaxy clusters are found from the SDSS DR5 database
in the redshift range $0.05<z<0.1$, of which 212 are new.
Local maximum-density galaxies successfully represent cluster centers.
We provide the cluster catalogue including a number of cluster parameters.
\end{abstract}


\section{Introduction}
The role of environment in the formation and evolution of
galaxies is a topic of much interest in modern astrophysics.
\citet{ab4} found that the abundance of early-type galaxies
increases in dense environment indicating that environment has
an effect on galaxy morphology. This \textit{morphology-density
relation} has been further studied by many authors
\citep[e.g.,][]{ab11,ad3,ac4,abb3,ac8} and has been extended
to lower density regions and high-redshift clusters
\citep{ae8,aba4,ac9}. There have been several physical
mechanisms proposed to explain this relation, such as ram
pressure stripping \citep{ad4,abb4,abc4,aa1,af8,ac3}, tidal
forces \citep{aba3,ad9}, galaxy harassment \citep{ac6,aba6},
starvation \citep{ae4}, and interaction and merging
\citep{aba9,aba5,ac2,ad2,ac3}.

Some of these processes involve violent gas dynamics, and hence
it is quite likely that environment influences the star formation
rate in galaxies \citep{abe4,af4,ag4,abf4,abc3,abd4}.
The cosmic star formation history itself is also suspected to
depend on environment \citep{abf9}.
Perhaps as hints of these suspicions, galaxy properties such as color,
luminosity, and size appear to depend on environment
\citep{aaa1,aab1,aca2,aca3,ad8,ag8,ac8}. However, the
exact physical processes that link between galaxy properties and
environment are still far from being clear.

Galaxy clusters are the most dense regions in the large-scale universe.
The number density of galaxies in clusters can be several hundred
times larger than that of the field. They are thus ideal
laboratories for studying the effect of environment on the formation and
evolution of galaxies.

The first galaxy cluster catalogues \citep{ac1,a12,ad1} were
based on the visual inspection of photographic plates. While
these catalogues have been widely adopted, they suffer from
many selection effects and spurious detections due to
projection effects. Furthermore, the visual inspection of large
parts of the sky is very time consuming. Systematic searchers
for galaxy clusters are needed to ensure the efficiency and
reliability of the cluster detection.
Among the most popular in the absence of spectroscopic information
were the technique using the early-type galaxy
color-magnitude relation \citep{abea4} and finding brightest cluster
galaxies (BCGs; \citealt{ae1,aba10}).
The first automatic search for optical clusters looking for
overdense regions was by \citet{abd9} followed by numerous others
\citep{ah4,acc3,ada8,acd3,adc4,adb4,ab6}.

With the advent of large-area, spectroscopic redshift surveys,
such as the Sloan Digital Sky Survey (SDSS; \citealt{ac11,abb9})
and Two Degree Field Galaxy
redshift Survey (2dFGRS; \citealt{acb3}), large
cosmological volumes with redshift have become
available. We are finally able to systematically search for galaxy
clusters in a large volume by measuring the local 3-d number density
of galaxies.
While the 3-d densities measured based on the spectroscopic database
are more powerful than projection-based 2-d searches for delineating
the local galaxy distribution, they require a high level of
coverage of member galaxies in the spectroscopic survey a priori.
However, even the most up-to-date surveys (such as the SDSS) show
only $\sim 60$--70\% completeness rate in dense regions, hampering
us from measuring the densities accurately.
In this paper, we introduce an improved method for finding galaxy clusters by
measuring local densities using both spectroscopic and
{\em photometric} data, and provide a new catalogue of the clusters we
have found.
We assume cosmological parameters $\Omega_m=0.3, \Omega_{\Lambda}=0.7,
q_0=-0.55$, and $H_0=70\ \rm km\ s^{-1}\ Mpc^{-1}$ throughout this paper,
and all distances are comoving.

\section[Data]{The Data}

The SDSS is performing a survey to cover a quarter of the whole sky.
The imaging survey of the SDSS DR5 contains 215 million objects
in 8000 $\rm deg^2$. The SDSS spectroscopic survey mapped
5740 $\rm deg^2$ obtaining about a million spectra of which
674,749 objects are classified as galaxies. The galaxy redshifts
provided  make it possible to study 3-d structure of galaxy distribution.

The SDSS photometric pipeline provides several kinds of
magnitudes. For galaxy colors, we use the \texttt{modelMags}
as they provide unbiased colors regardless of any color
gradients. For galaxy luminosities, we use
\texttt{petroMags} as these are a better estimate of the total
luminosity \citep{abb9}. We also apply a K-correction as described
in \citet{ae2} and correct for Galactic extinction with the
\citet{abg9} values provided by the SDSS pipeline.

We use the SDSS DR5 both photometric and spectroscopic data. For the
galaxies with spectra, we extract {\em all types of galaxies} in the range
$0.05<z<0.1$ and having $r_{\rm petro}<17.77$ \citep{abc9}.
At $z = 0.1$, $r = 17.77$ corresponds to an absolute magnitude of
$M_r = -20.55$. In order to create a volume-limited sample and
so avoid biases with redshift, we cut at this absolute magnitude.

For galaxies without spectra, we select all galaxies with
$13.00<r_{\rm petro}<17.77$ to have the same
apparent magnitude cut as the spectroscopic data.
Its faint limit comes from the SDSS spectroscopic survey limit,
and the bright limit is from the fact that the objects of
$r_{\rm petro}<13.00$ and $z>0.05$ are almost always stars
rather than galaxies.

\section[Measuring Density]{Measuring Galaxy Density}

\subsection{Local Density Measurement}

Popular methods to measure galaxy local densities include
a simple estimation of spatial number density in a certain radius
and the distance to the {\it n}th nearest galaxy.
The number density, however, cannot tell us about the concentration
status which is critical for studying dynamical evolution, while
using the {\it n}th nearest galaxy can be biased by local density
fluctuations. Thus, in addition to the number density,
the spatial separation between galaxies can be taken into account
as well to provide additional information.
For example, closer neighbors can be
weighted more than distant galaxies. \cite{ab9} (hereafter S06)
introduced such a weighting scheme with a Gaussian filter.
The S06 scheme considers neighboring galaxies in the ellipsoid defined by:
\begin{equation}
  {\left(\frac{r_{a}}{3 \sigma}\right)^2 + \left(\frac{r_{z}}{3 c_{z} \sigma}\right)^2 \leq 1},
\end{equation}
where $\sigma$ is the searching distance criterion, $r_a$ is the projected
distance and $r_z$ is the line-of-sight
distance, all in Mpc, to a neighboring galaxy.
It allows all the galaxies within the ellipsoid to be counted
for measuring the density.
The $c_z$ factor is a simple compensation for
``the finger-of-god'' effect due to the peculiar motion
and is estimated by counting the number of
galaxies $n$ within the sphere of radius $\sigma$ as:
\begin{equation}
 c_z = 1 + 0.2 n
\end{equation}
where $n$ is capped at 10. Hence, this can scale along the radial
direction up to a factor of 3. Density measures by adopting a fixed
volume in comoving space can lead to sampling vastly different
volumes in the field and in clusters due to the stretching of
the line of sight direction by the finger-of-god effect.

The density parameter to each member galaxy based on the
spectroscopic data is calculated as
\begin{equation}
  \rho_{\rm spec,3D}(\sigma) = \frac{1}{\sqrt{2 \pi}\sigma } \sum_i \exp \left[ -\frac{1}{2} \left(\frac{r_{a,i}^2}{\sigma^2} + \frac{r_{z,i}^2}{c_{z}^2 \sigma^2} \right) \right]
  \label{rho_g2}
\end{equation}
by summing up for all the members within the search ellipsoid.
The target galaxy is not included in the density summation.
Our algorithm estimates the local density with surrounding galaxies
in a relatively large area and thus minimizes local effects.

In this paper, we attempt to improve the density measurement scheme
of S06 by adding photometric member candidates and build on the S06 algorithm
to design an effective cluster search method.

\subsection{Member Selection with Spectroscopic Data}

Since we do not have the line-of-sight information on the photometric
member galaxies, we first measure the 2-d densities for our
spectroscopic galaxy samples ignoring the line-of-sight information.

We first define an ellipsoid as in S06:
\begin{equation}
  {(\frac{r_{a}}{\sigma_1})^2 + (\frac{r_{z}}{\sigma_2})^2 \leq 1}.
\end{equation}
We choose $\sigma_1 = 1$\,Mpc because it optimizes the search for the
density peak and the brightest cluster galaxy of the cluster.
For $\sigma_2$ we adopt the distance corresponding to
three times the velocity dispersion of the cluster.
Note that $\sigma_2$ is used only for selecting member galaxies but
not for the 2-D density measurement.
Readers are referred to \S 3.5 for more details.
Galaxies within the ellipsoid are regarded as member galaxies of
the cluster.
Then the new 2-d density $\rho_{\rm spec}$ for the spectroscopic members is:

\begin{equation}
  \rho_{\rm spec} = \frac{1}{\sqrt{2 \pi}\sigma_1 } \sum_{i} \exp \left[ -\frac{1}{2} \left(\frac{r_{a,i}^2}{\sigma_1^2}\right) \right]
  \label{rho_g1}
\end{equation}
where $\sigma_1 = 1$\,Mpc.

\subsection{Member Selection with Photometric Data}

\begin{figure}
 \begin{center}
  \vspace{5pt}
   \includegraphics[width=\columnwidth]{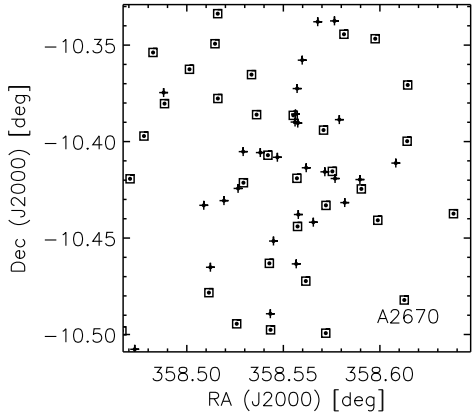}
   \caption{The spectroscopic completeness for the case of Abell 2670
with $r_{\rm petro} < 17.77$.
The dots with rectangles are galaxies covered by the SDSS
spectroscopic survey and the crosses are the missed ones.
The completeness rate for this cluster is only 55\% in
this field of view and 65\% within 1\,Mpc radius from the cluster center.
This figure illustrates the problem of fiber collisions in dense environments.}
 \label{A2670}
 \end{center}
\end{figure}

The SDSS spectroscopic survey tries to cover galaxies as
completely as possible with a tiling algorithm \citep{af2}.
However, its selection scheme can leave some galaxies unobserved
due to fiber collisions.
Figure \ref{A2670} shows the extreme case of Abell\,2670 which shows
only 65\% of spectroscopic coverage rate within the 1\,Mpc radius
from the cluster center.
This effect is generally worse for denser regions because the fiber
collision problem is obviously worse in the more crowded regions.
We estimate the spectroscopic completeness, that is, the number
fraction of galaxies covered by the SDSS DR5 spectroscopic survey,
to be $f_{\rm spec} \sim 65\%$ for rich clusters. An independent study
has also reported that $30\%$ of true brightest cluster galaxies are
missed by the spectroscopic survey \citep{aba10}.
The ``incompleteness'' of the spectroscopic survey causes
a problem in measuring the densities of galaxies in dense regions.
We attempt to alleviate this by further considering photometric data.

\begin{figure}
 \begin{center}
  \vspace{5pt}
  \includegraphics[width=\columnwidth]{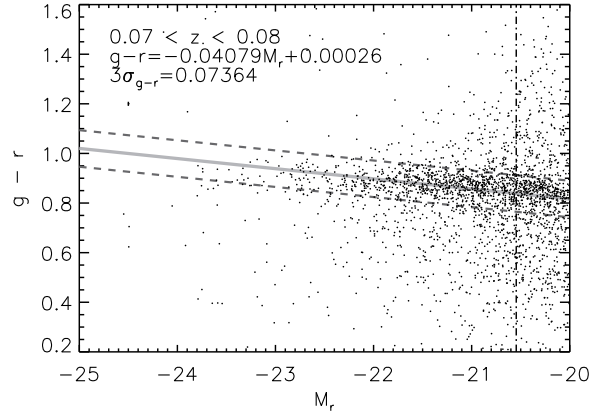}
  \caption{
Empirical CMR for the redshift range $0.07 < z < 0.08$. The gray-solid
line shows a linear fit after $2\sigma$ clipping until the number of galaxies
on the CMR becomes constant and the dashed lines show the scatter
(the gray-solid line $\pm3\sigma$ in $g-r$ color).
The vertical dot-dashed line is a absolute magnitude cut in our data
selection.}
 \label{CMR}
 \end{center}
\end{figure}

Early-type galaxies in clusters have a tight correlation
between their optical colors and luminosity, known as the
color-magnitude relation (CMR). The CMR was first observed by
\citet{ab1} and is often reported to be universal within errors
\citep{ab10,ak4,al4}. Based on these assumptions, we can select
candidate cluster members even though they do not have redshift
because galaxies on the CMR have a
high likelihood of belonging to the cluster. Thus, we find the
slope and scatter of the CMR in order to select
``photometric members''. For small clusters, the slope of the CMR can
be difficult to achieve. Hence, we construct an empirical CMR by
stacking galaxies that are members of 20 typical clusters
determined by their redshifts.
We do this for five redshift bins with $\Delta z = 0.01$, and an
example is shown in Figure~\ref{CMR}. A linear
fit with $2\,\sigma$ clipping is iterated until the number of
galaxies on the CMR remains constant. The galaxies residing
on the CMR $\pm (3 \times rms)$ in $g-r$ color are
selected as ``photometric members''.
We compute the 2-d photometric density $\rho_{\rm phot}$ following
Eq. \ref{rho_g1} using only the photometric members.

The total density $\rho$ of a galaxy is calculated by combining the
spectroscopic and photometric density parameters:
\begin{equation}
  \rho = \rho_{\rm spec} + \rho_{\rm phot}.
\end{equation}
In the process of combining $\rho_{\rm spec}$ and $\rho_{\rm phot}$
we are losing the line-of-sight information and our density measures
are projected into only two dimensions; but, our method still
provides improved density measures as will be demonstrated
in \S 4. through our follow-up spectroscopic observations.

 When we determine whether a particular galaxy without redshift
belongs to a cluster as described in this section, we
assume that the galaxy is at the redshift of the cluster and
apply the same absolute magnitude cut as discussed in \S 2.

We have attempted improving our local density measure further by
applying a galaxy luminosity weight (i.e. weighting more luminous
galaxies more) but found no significant difference. Hence,
we have decided to ignore it.

\subsection{The Efficiency of the CMR technique}

We expect that the CMR to be effective for finding cluster member
galaxies but that it may still suffer from the projection effect of
background and foreground objects. Hence, we test its efficiency.
We restrict the test on the spectroscopic data only, because only
with redshift we can determine whether a particular galaxy is a cluster
member or not, that is, for a convincingly high confidence.
%We use the "true" members determined by their redshift. The CMR
%can find red members only as mentioned in Section 3.4. Thus, we do not
%take into account blue member galaxies in this test.
Besides, we use early-type galaxies alone for this test as our CMR
method works only for them.
In order to select early-type galaxies (ETGs), we use the
morphological index \textit{fracDev} from the SDSS pipeline, which
shows the weight of the de Vaucouleurs profile in the two-component
profile fit with the deVaucouleurs and exponential profiles.
We apply the highly conservative $r$-band $fracDev > 0.95$ cut,
and the galaxies selected are highly likely to be early type.
Using the spectroscopic members meeting the same $fracDev$
criterion, we check the validity of our scheme defining
the following two CMR-related quantities.
\begin{eqnarray}
\mbox{Completeness} &=& {\textrm{\# of ETG members in CMR} \over
                  \textrm{\# of all ETG cluster members}} \times 100 \\
\mbox{Purity} &=& {\textrm{\# of ETG members in CMR} \over
            \textrm{\# of ETG galaxies in CMR}} \times 100
\end{eqnarray}

\begin{figure}
 \begin{center}
  \vspace{5pt}
  \includegraphics[width=\columnwidth]{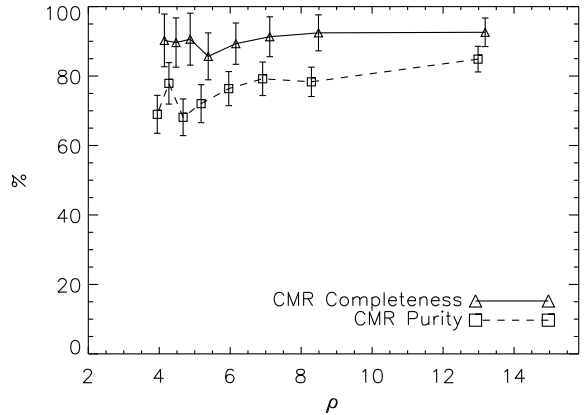}
  \caption{
The solid and dashed lines show completeness and purity of
the photometric member candidates found from our technique
(see the text for definition). The errors are from
the Poisson statistics in the equal number bin. The rectangles are shifted
slightly to the left for the clarity of the figure.
The completeness is constantly $\sim 90\%$ and the purity is 70--85\%
with an apparent density dependence.  The high values of these
parameters validate our CMR technique.}
 \label{cmr_test}
 \end{center}
\end{figure}

The ``CMR completeness'' is a measure of the number of the early-type
cluster members that also sit on the CMR.
The ``CMR purity'' tells us the fraction of galaxies selected by
the CMR method that are indeed early-type cluster members.
Figure~\ref{cmr_test} presents the CMR completeness and purity
of our clusters as a function of density.
The CMR completeness is around 90\% meaning that 90\% of the
early-type (by $fracDev$) member galaxies indeed sit on the CMR.
The purity is 70--85\% meaning that only 70--85\% of the galaxies
on the CMR are early-type (by $fracDev$) members. The rest may be
background or foreground galaxies. Exact values of these parameters
modestly depend on the $fracDev$ criterion adopted, but on the whole
both parameters show high values validating our scheme.

\subsection{Finding Clusters}

The SDSS DR5 photometric data includes about a million galaxies in
the magnitude range described in \S 2. Searching for clusters through
the entire data set would require an exhaustive amount of effort.
So we begin our cluster search first by estimating the local density
of $spectroscopic$ members and identify about 7000 galaxies with high density
measures as candidate cluster positions. For this selection
we use 3-D spectroscopic density parameter ($\rho_{\rm spec,3D} > 0.5$).

We then select all the photometric and spectroscopic galaxies within a 2
Mpc radius from the cluster center candidate galaxies and
estimate their total 2-d local density with both the spectroscopic
and photometric members for all galaxies in this radius.

With these 2-d spectro-photometric density measurements,
we find the galaxy with the
highest density in the area of a 2 Mpc circle and $\Delta z=0.01$. The
highest density galaxy in a local area is defined as the
maximum-density galaxy (hereafter MDG). As the MDG resides at the
densest part of the selected volume, we define it as the
cluster center and declare it a cluster if the $\rho$ of an
MDG is greater than 4.0 which roughly corresponds to a
cluster velocity dispersion of $200\, \rm km\,s^{-1}$.

We $expect$ that the MDG should be close to the cluster
center and coincide with the brightest cluster galaxy
(BCG; \citealt{ab5,ab8}).
We compared the projected separation of BCGs to MDGs and
found that they do not always coincide, however.
We assume that a good cluster search scheme should
identify MDGs as BCGs as well, and thus we vary the
member search radius (the short axis of the ellipsoid)
hoping to minimize the separation.

\begin{figure}
 \begin{center}
  \vspace{5pt}
   \includegraphics[width=\columnwidth]{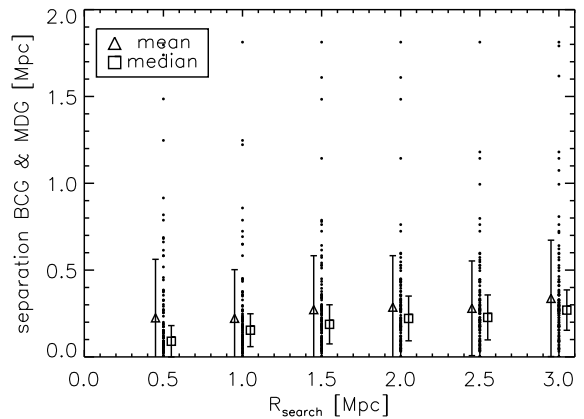}
   \caption{
The projected separation between brightest cluster galaxies (BCGs) and
maximum-density galaxies (MDGs). Triangles are the mean and squares are
the median separations. The error bars show the standard deviation
for the mean and the median absolute deviation for the median.
When the search radius to find
member galaxies is 1\,Mpc, the separation is small enough that
the cluster search is effective. But a substantially smaller values than
1\,Mpc would miss many memeber galaxies because the typical size of clusters
can be easily as large as 1\,Mpc (see the text).}
 \label{BCG_separation}
 \end{center}
\end{figure}

Figure~\ref{BCG_separation} shows the separation between MDGs and BCGs.
The mean and error bars reflect the large scatter in the sample.
The median is a better estimate of the typical distance and
shows a convergence towards a lower value of search radius.
%In Figure~\ref{BCG_separation}, the separation between MDGs and BCGs
%decreases as the search radius gets smaller.
But too small a value of
the search radius would miss too many member galaxies in the density
measurement and be dominated by local density fluctuations.
The typical Virial radius of our clusters is 0.5\,Mpc in
the 2-d projection, and a substantial fraction ($\sim 50$\%)
of clusters have a Virial radius greater than 0.5\,Mpc and extend
beyond 1\,Mpc. Considering this, we have chosen 1\,Mpc as our search radius.
It should be noted that our search using spectroscopic data and the
CMR is liable to
missing clusters that are dominated by blue, starforming galaxies, if any.

\section[Result]{Results}

\subsection{Cluster Catalogue and Properties}

\begin{table*}
 \centering
  \caption{Cluster catalogue with $\rho \geq 4$ }
  \begin{tabular}{ccccccccccccc}
  \hline
   (1) & (2) & (3) & (4) & (5) & (6) & (7) & (8) & (9) & (10) & (11) & (12) \\
   ID & $\rm ra_{MDG}$ & $\rm dec_{MDG}$ & $\rm z_{clt}$ & $\rho$ & $\rm N_{200}$ & $\rm R_{200}$ & $\sigma_{\rm v}$ & $\rm ra_{BCG}$ & $\rm dec_{BCG}$ & $\rm z_{BCG}$ & Comments \\
    & (J2000) & (J2000) & & & & [Mpc] & [$\rm km\,s^{-1}$] & (J2000) & (J2000) & &\\
 \hline
   1 & 258.20851 &  64.05295 & 0.08257 & 32.197 & 148 & 2.785 & 1299 & 258.12006 &  64.06076 & 0.07340 &               A2255 \\
   2 & 358.55746 & -10.39042 & 0.07623 & 29.234 &  91 & 1.842 &  840 & 358.55703 & -10.41904 & 0.07766 &               A2670 \\
   3 & 208.26551 &   5.13824 & 0.07945 & 25.020 &  50 & 1.421 &  780 & 208.27667 &   5.14974 & 0.07890 &               A1809 \\
   4 & 228.81884 &   4.37958 & 0.09796 & 24.611 &  65 & 1.487 &  864 & 228.80879 &   4.38621 & 0.00000 &               A2048 \\
   5 & 255.66908 &  33.52032 & 0.08819 & 22.783 &  74 & 1.813 & 1135 & 255.63809 &  33.51666 & 0.08640 &               A2245 \\
   6 & 239.58334 &  27.23342 & 0.08969 & 22.476 & 190 & 2.918 & 1000 & 239.58334 &  27.23342 & 0.09081 &               A2142 \\
   7 & 205.54886 &   2.22449 & 0.07688 & 21.140 &  41 & 1.286 &  845 & 205.54018 &   2.22721 & 0.00000 &               A1773 \\
   8 & 186.91015 &   8.82133 & 0.08981 & 20.599 &  41 & 1.305 &  849 & 186.87809 &   8.82456 & 0.00000 &               A1541 \\
   9 & 230.33576 &  30.67093 & 0.07845 & 19.313 &  76 & 1.723 &  677 & 230.33576 &  30.67093 & 0.07845 &               A2061 \\
  10 & 255.64172 &  34.07809 & 0.09871 & 19.178 &  64 & 1.546 & 1120 & 255.67708 &  34.06003 & 0.09891 &               A2244 \\
 \hline
\label{catalogue}
\end{tabular}
\tablecomments{
Column 1: ID;
Column 2,3: the coordinates of the maximum-density galaxies (MDGs);
Column 4: the redshift of clusters estimated with spectroscopic data;
Column 5: the density of the cluster MDG within 1\,Mpc;
Column 6: the number of member galaxies of clusters within $R_{\rm 200}$;
Column 7: the virial radius of clusters obtained by the number density
profile of the galaxy clusters;
Column 8: the velocity dispersion within $R_{\rm 200}$;
Column 9,10: the coordinates of the brightest-cluster galaxies (BCGs);
Column 11: the redshift of the BCGs. The entries with zero mean the
absence of the redshift information;
Column 12: matching result with known catalogue of galaxy clusters.
The complete version of the galaxy clusters is available on the web-page
http://gem.yonsei.ac.kr/html/cluster.php
}
\end{table*}

\begin{figure*}
 \begin{center}
  \vspace{5pt}
  \includegraphics[width=1.5\columnwidth]{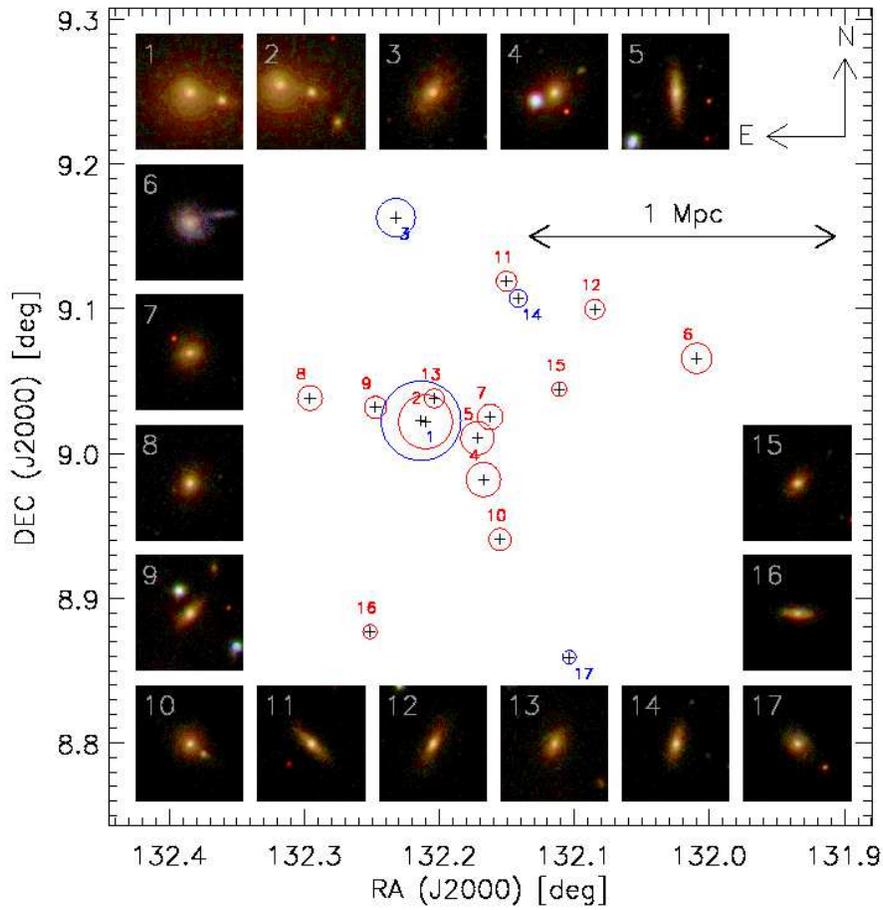}
  \caption{A sample rich cluster that is newly found in our study.
The density ranking is 142th, out of 924, with  the density parameter
$\rho = 9.1$ and $N_{\rm gal} = 16$ within 1\,Mpc, where its virial
radius $R_{\rm 200}$ is estimated to be 0.737\,Mpc and $N_{\rm 200} =11$.
The size of each circle denotes its $r$-band luminosity. Blue and red circles
show photometric and spectroscopic members, respectively. SDSS image cut-outs
are shown for the 16 members and the central MDG (number 7). The BCG is
marked as number 1.}
 \label{new_clt}
 \end{center}
\end{figure*}

We provide our cluster catalogue in Table \ref{catalogue}.
We have found 924 galaxy cluster candidates with $\rho \geq 4.0$
which roughly corresponds to $200\,\rm km\,s^{-1}$.
Among them, 212 clusters are newly found in this study.

Figure~\ref{new_clt} shows a sample from the newly-found clusters.
The BCG and MDG are marked as number 1 and 7, respectively.
Blue and red circles denote photometric and spectroscopic members,
while the size of circles is proportional to the galaxy $r$-band brightness.
Postage images of the member galaxies are also shown.
Its MDG is ranked 142th out of 924, with
$\rho = 9.1$ and $N_{\rm gal} = 16$ within 1\,Mpc.
For comparison, the central galaxy in the Virgo cluster, M87, has
$\rho = 7.5$ and $N_{\rm gal} = 19$ within 1\,Mpc, and thus is comparable.

Table \ref{catalogue} provides various properties of our galaxy clusters.
The positions of MDGs and BCGs are included.
The redshift of each cluster is obtained by finding the redshift peak
of the spectroscopic members and fitting it by Gaussian function.
When the number of spectroscopic members is less than 4, we simply find median
redshift of spectroscopic member galaxies.
We calculate the virial radius $r_{\rm 200}$ by counting the number
density of member galaxies and comparing it to the cosmological
background number density of galaxies. We estimate the number of
galaxies within a virial radius $N_{\rm 200}$ by counting the galaxies
within $r_{\rm 200}$.
The velocity dispersion is calculated with biweight
estimator \citep{ab2} in $r_{\rm 200}$ after 3-sigma clipping.
As a proxy to the cluster size and density, we can use $\rho$
regardless of the cluster size (virial radius).

\begin{figure}
 \begin{center}
  \vspace{5pt}
  \includegraphics[width=\columnwidth]{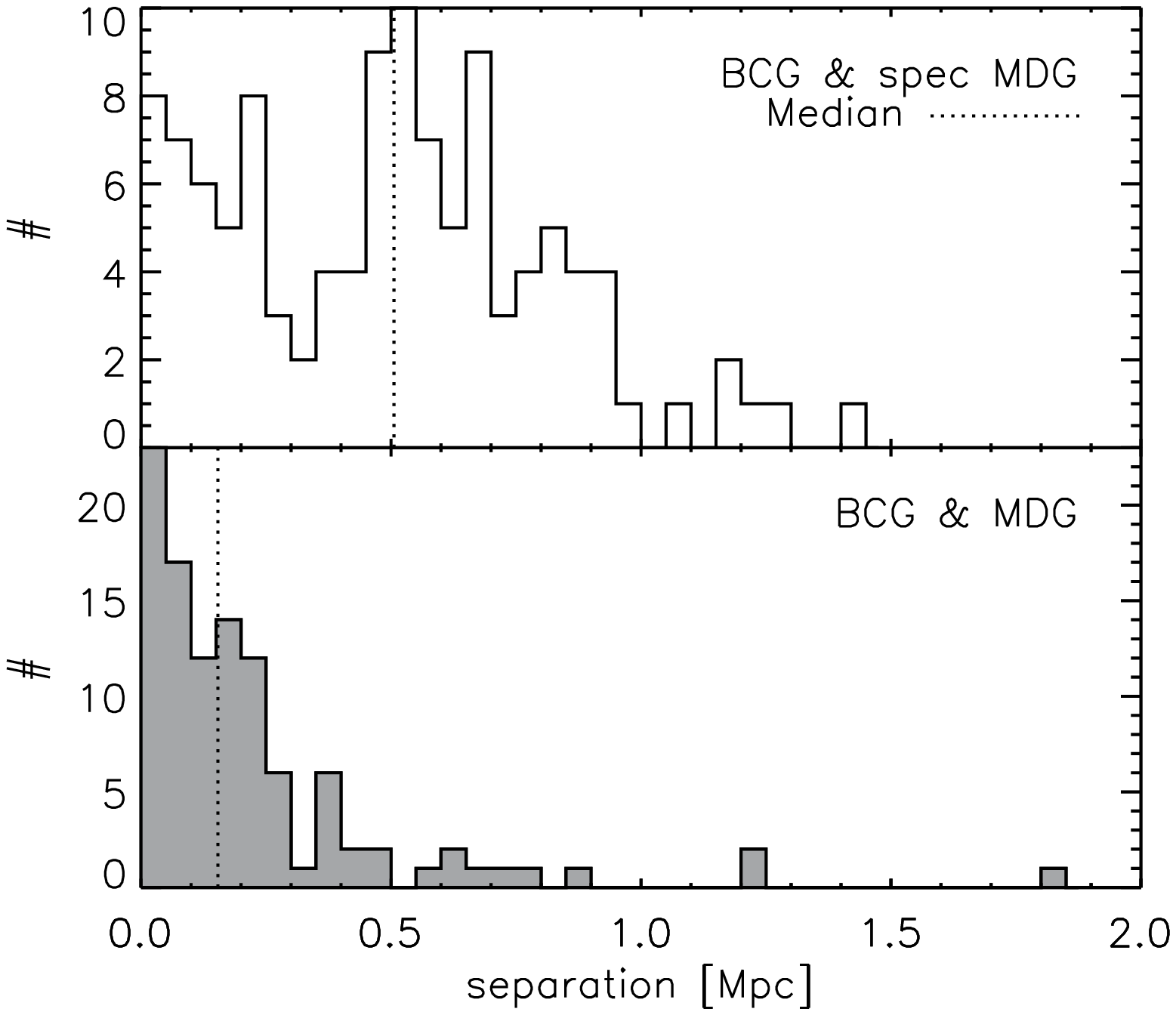}
  \caption{
We show the separation between BCGs and MDGs. The dotted lines
represent the median separation. The top panel shows
the separation when the CMR-selected photometric members are not included
in the density measurement. When we add the photometric members
(bottom), the separation gets much smaller yielding a more
reliable cluster center detection.}
 \label{separation_hist}
 \end{center}
\end{figure}

In \S 3.5, we introduced the separation between the MDGs
and BCGs as a test of how well the cluster search scheme finds
cluster centers. In Figure~\ref{separation_hist}, we show
the separation between MDGs and BCGs in our method with (bottom)
and without (top) added photometric members. It illustrates
the effectiveness of the added photometric members.

\begin{figure}
 \begin{center}
  \vspace{5pt}
  \includegraphics[width=\columnwidth]{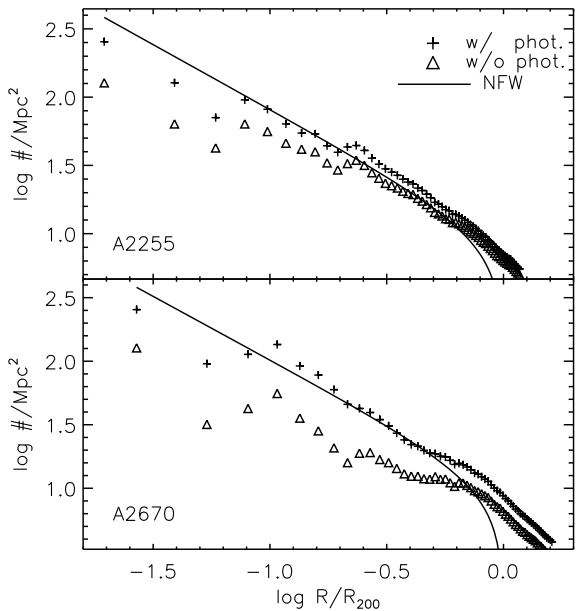}
  \caption{
The projected radial profiles of the galaxy number density of
sample rich clusters centered at their BCGs. The plus signs show the galaxy
number density profiles from our scheme (including both spectroscopic and
photometric members), and the triangles are those with only
spectroscopic members. Also shown for comparison are the projected NFW
halo profiles fitted to our measurements $with$ both spectroscopic
and photometric members.
Our measurements are in good agreement with the projected NFW profiles.
}
 \label{profile}
 \end{center}
\end{figure}

The galaxy number density profile is a good proxy for the mass
profile and thus dark matter profile of galaxy clusters
\citep{abd3,ae9,add4,ai4,aj4}.
In Figure~\ref{profile}, we show the galaxy number density profile
of two examples. We compare them with the Navarro-Frenk-White (NFW)
dark matter profile \citep{ab7}, which is defined as
\begin{equation}
{{\rho(r)}\over{\rho_c}} = {{\delta_c}\over{(r/r_s)(1+r/r_s)^2}},
\end{equation}
and can be projected into a 2-dimensional space as
\begin{equation}
  {{\rho(r_p)}\over{\rho_c}} = {2 n_0 \int_0^\theta {{\delta_c}\over
    {(r_p/r_s cos\theta)(1+r_p/r_s cos\theta)^2}}d\theta}
\end{equation}
where $n_0$ is normalization factor, $r_p$ is projected distance and
${\theta}={cos^{-1}{{r_p}\over{r_{\rm 200}}}}$.
The number density profile is in reasonable agreement with the
NFW profile as found also by \citet{aj4}.
The galaxy number density profile only based on the spectroscopic members
is lower than the one with photometric members added. The departure gets
larger in the central region of the clusters where the spectroscopic
coverage is poor.

\begin{figure*}
 \begin{center}
  \vspace{5pt}
  \includegraphics[angle=90,width=2.2\columnwidth]{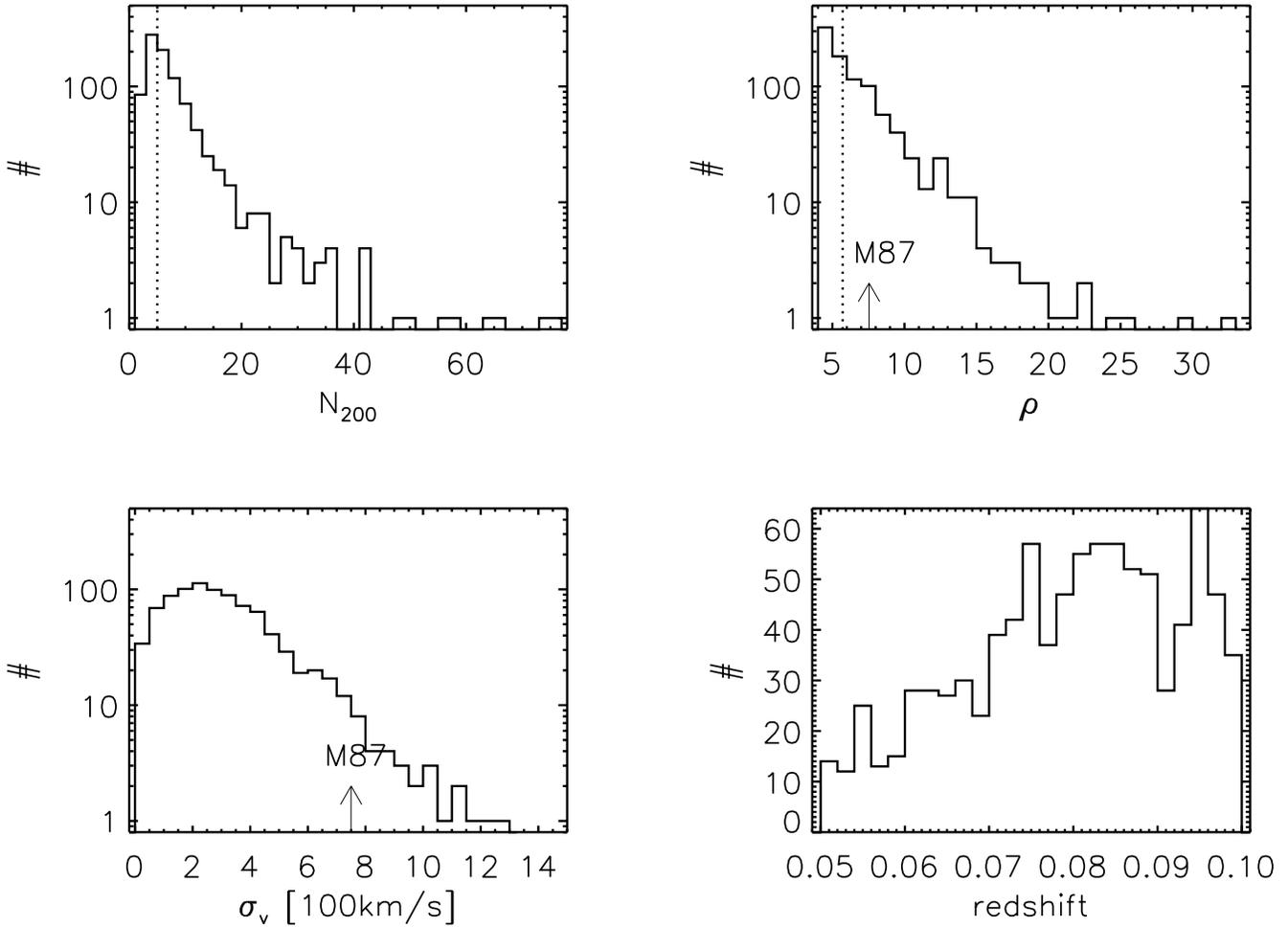}
  \caption{
Histogram of $N_{\rm 200}$, $\rho$, velocity dispersion, and redshift.
We define galaxy clusters when a condition $\rho \geq 4.0$ is met.
In the top left panel two clusters of $N_{\rm 200} = 190$ and 148 are
out of bound. The vertical dotted lines show median values.
The density and velocity dispersion of M87 are shown as the arrow.}
 \label{histogram}
 \end{center}
\end{figure*}

A larger cluster would have a higher value of $\rho$, the density measure
within 1\,Mpc.
We show the distribution of $\rho$, $N_{\rm 200}$ and the velocity
dispersion of clusters within $r_{\rm 200}$ in Figure~\ref{histogram}.
The (minimum, median, and maximum) for the four panels are
%corrected!!
%(0, 6, 180) for $N_{200}$, (4.0, 5.7, 32.2) for $\rho$,
%(0, 261, 1315) for $\sigma$, and the redshift range is 0.05 through 0.1.
(0, 5, 190) for $N_{200}$, (4.0, 5.7, 32.2) for $\rho$,
(0, 263, 1299) for $\sigma$, and the redshift range is 0.05 through 0.1.
The density and the velocity dispersion of M87 \citep{Binggeli}
in Virgo cluster is shown as well.

\begin{figure}
 \begin{center}
  \vspace{5pt}
  \includegraphics[width=\columnwidth]{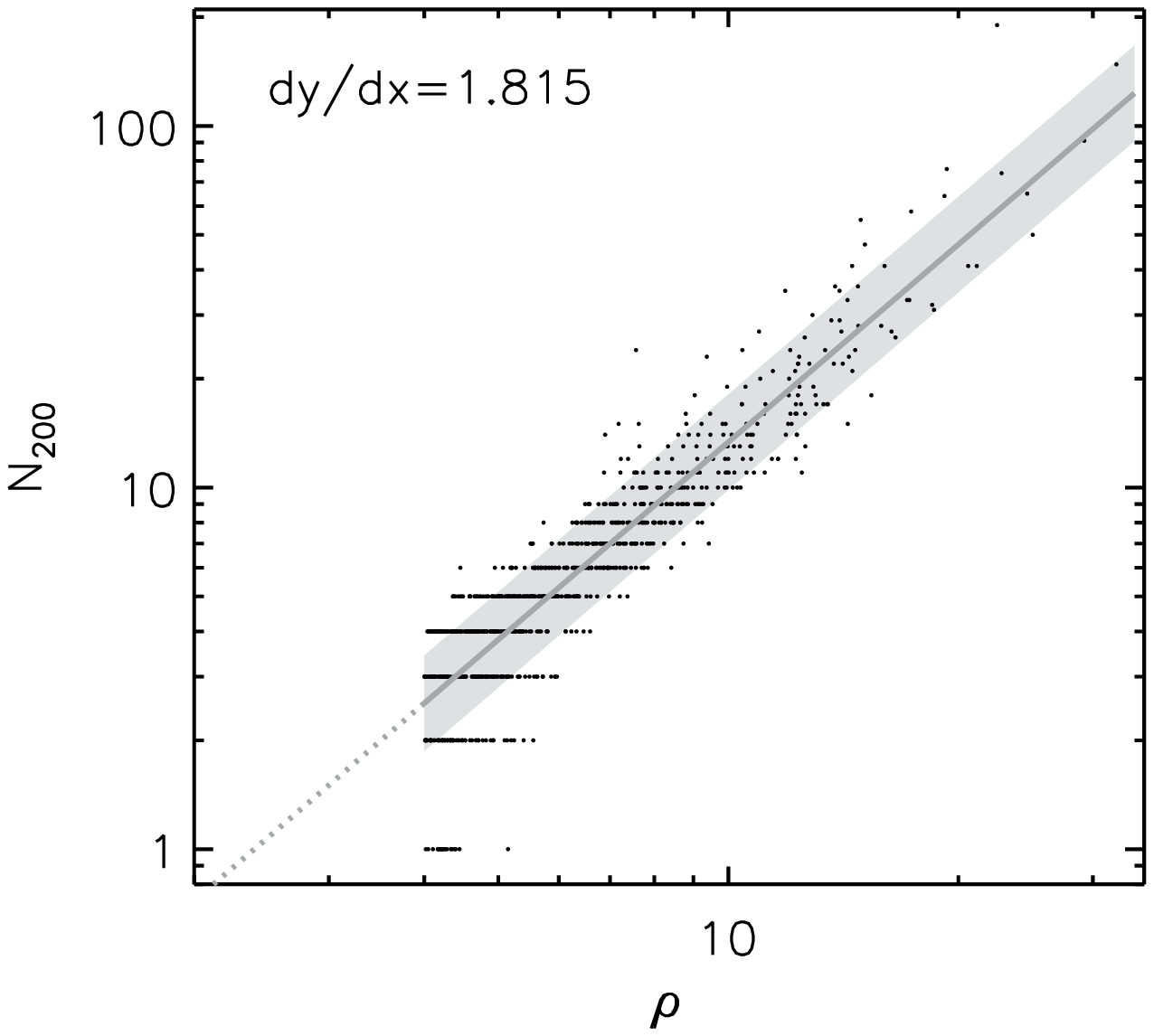}
  \caption{
The density parameter $\rho$ is proportional to $N_{\rm 200}$ scaling
as power law. Note that the density parameter includes the distance
information in addition to the number. The thick line is the linear
fit to the data and the grey band shows the standard error of the fit.}
 \label{rho_N}
 \end{center}
\end{figure}

In Figure \ref{rho_N} we compare $\rho$ with $N_{\rm 200}$.
Also shown is the linear fit to the data.
The density parameter $\rho$ appears to scale
as a power law with $N_{\rm 200}$ rather than with a linear relation.
That is, in dense regions, the linear fit would have a steeper
slope. This is because the Gaussian weight in our density
measuring scheme takes the distance to each member into account.
For a given number of cluster members, a more concentrated cluster
would have a higher value of $\rho$.

\begin{figure}
 \begin{center}
  \vspace{5pt}
  \includegraphics[width=\columnwidth]{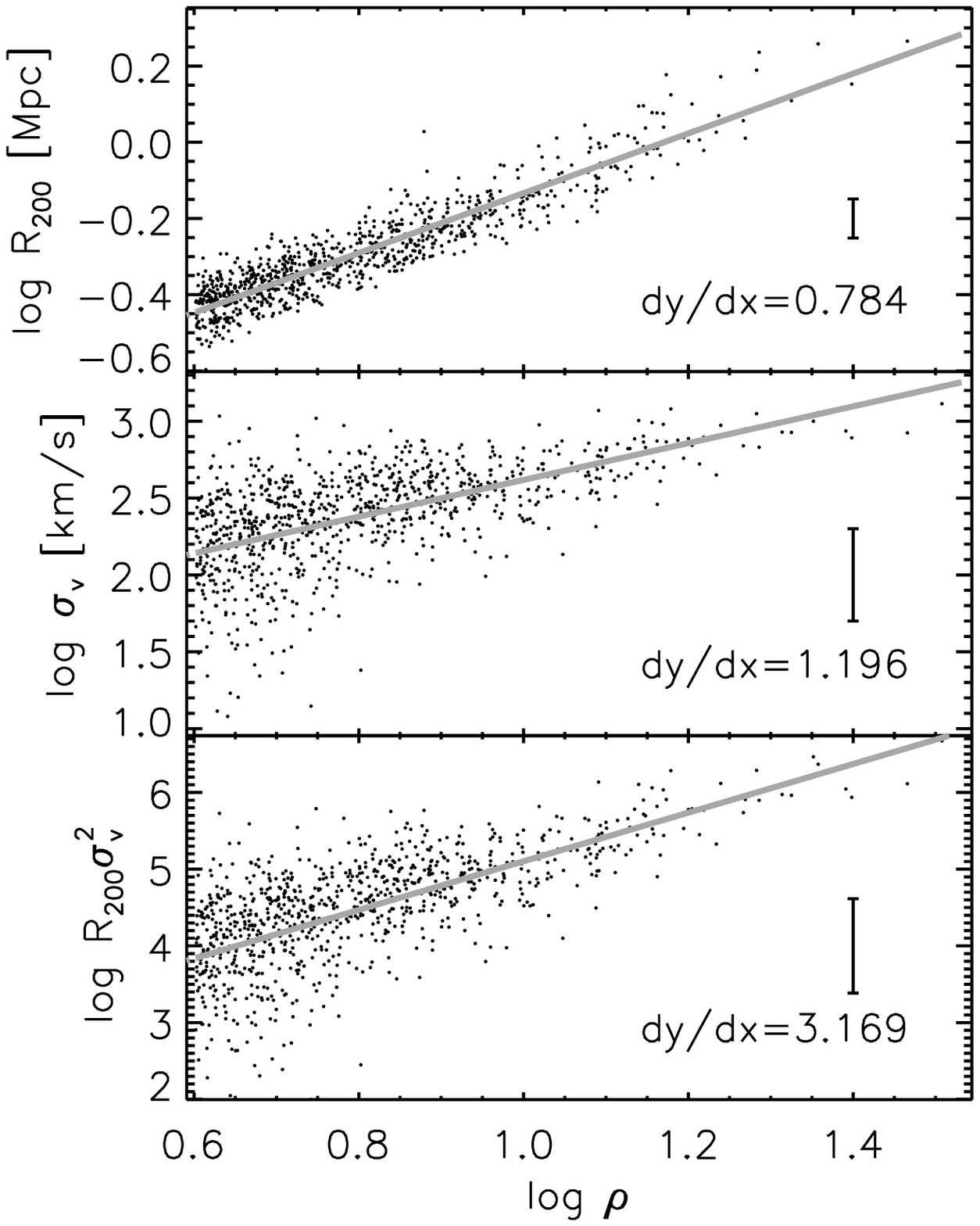}
  \caption{The density parameter $\rho$ is in agreement
    with the properties representing the mass and size of clusters.
    The virial radius in the top panels has a tight correlation with $\rho$.
    The density parameter correlates with the velocity dispersion and
    dynamical mass in the middle and bottom panels.
    The thick gray line is our linear fit and $dy/dx$ presents
    the exponents of the power law. The error bars represent
    mean scatter.}
 \label{rhoN_r}
 \end{center}
\end{figure}

In Figure \ref{rhoN_r} we compare between various cluster properties.
The density parameter is a good proxy to
the size of galaxy clusters. It also shows a good correlation with
the velocity dispersion of clusters. Thus, our density parameter
can be an obvious indicator of dynamical mass of a galaxy cluster as well,
as shown in the bottom panel.

\subsection{Comparison to other catalogue}

\begin{table}
 \centering
 \begin{minipage}{\columnwidth}
  \centering \caption{The number of galaxy clusters found in our scheme matched to
    previous catalogues.}
  \begin{tabular}{ccc}
  \hline
   Cluster Catalogue & Number & Recovery(\%) \\
  \hline
    Abell         & 128  & 79\% \\
    Zwicky        & 457  &      \\
    C4            & 300  & 66\% \\
    Others        & 889  & \\
    Newly Found   & 212  & \\
    Total         & 924  & \\
  \hline
\label{recovery}
\end{tabular}
 \end{minipage}
\begin{minipage}{\columnwidth}
\tablecomments{The existing cluster catalogues are from the
VizieR Catalogue Service.}
 \end{minipage}
\end{table}
We perform a cross-matching between existing catalogues and ours.
The matching criteria are 2\,Mpc in the projected radius and
$\Delta z=0.01$. When the known clusters do not have redshift
(e.g., Zwicky clusters),
we just match them within 2\,Mpc radius in the sky. Each cluster
in our catalogue is matched with other catalogues separately.
The clusters that are detected in our method are generally in the
existing catalogues (See Table \ref{recovery}). Most rich
clusters are matched with Abell or Zwicky clusters and this
implies that the new algorithm using MDGs successfully finds
rich galaxy clusters. For those clusters in the existing
catalogues which include redshift, we can check the rate at
which our method recovers clusters known previously. Some Abell
clusters have redshifts and the C4 catalogue clusters always
do \citep{ab6}. We cross-check our clusters against the Abell
and C4 clusters that are in the SDSS area and in our redshift
range. 128 out of 162 Abell clusters with
redshifts and 300 out of 458 C4 clusters are detected in our
method, corresponding to detection rates of 79\% and 66\%,
respectively.
If we lower the $\rho$ cut for ``clusters'' from 4.0 to 1.0
we would find 89\% and 95\% of Abell and C4 clusters, respectively.
For example, by using a $\rho=1$ cut, we can recover additional 18
Abell clusters. However, such a low value of cut ($\rho \sim 1.0$) would
also find significantly more cluster candidates (3,959) many of which
may be false detection.
Our density measures with
$\rho=4$ cut are in general based on about 7 member galaxies.
Besides, when only a small number of galaxies are used for density
measurement, the density measures become less reliable, too.
Since our goal is to construct a database of galaxy clusters with
$improved$ and hence reliable density measures, we think our
$\rho = 4.0$ cut is a good compromise.
However, we admit that this choice is still somewhat arbitrary
and by using this cut we are neglecting smaller clusters.
In this regard, our 924 cluster candidates represent a robust
sample of relatively-rich clusters.

The galaxy clusters that were known to Abell but missed by our
method are mostly due to the $\rho$ cut as mentioned above.
In addition, our method missed 9 Abell clusters which are mainly
composed of relatively faint galaxies that do not make it to
our volume-limited sample.
Two Abell clusters were missed because they are found by
our method to be parts of larger clusters, just like the M87 and M49
groups in Virgo cluster. Five additional clusters were missed
due to various minor reasons: e.g., mismatch in the redshift of the
center galaxy of a cluster, mismatch in the position of the center
galaxy of a cluster, and so on.

\begin{figure}
 \begin{center}
  \vspace{5pt}
  \includegraphics[width=\columnwidth]{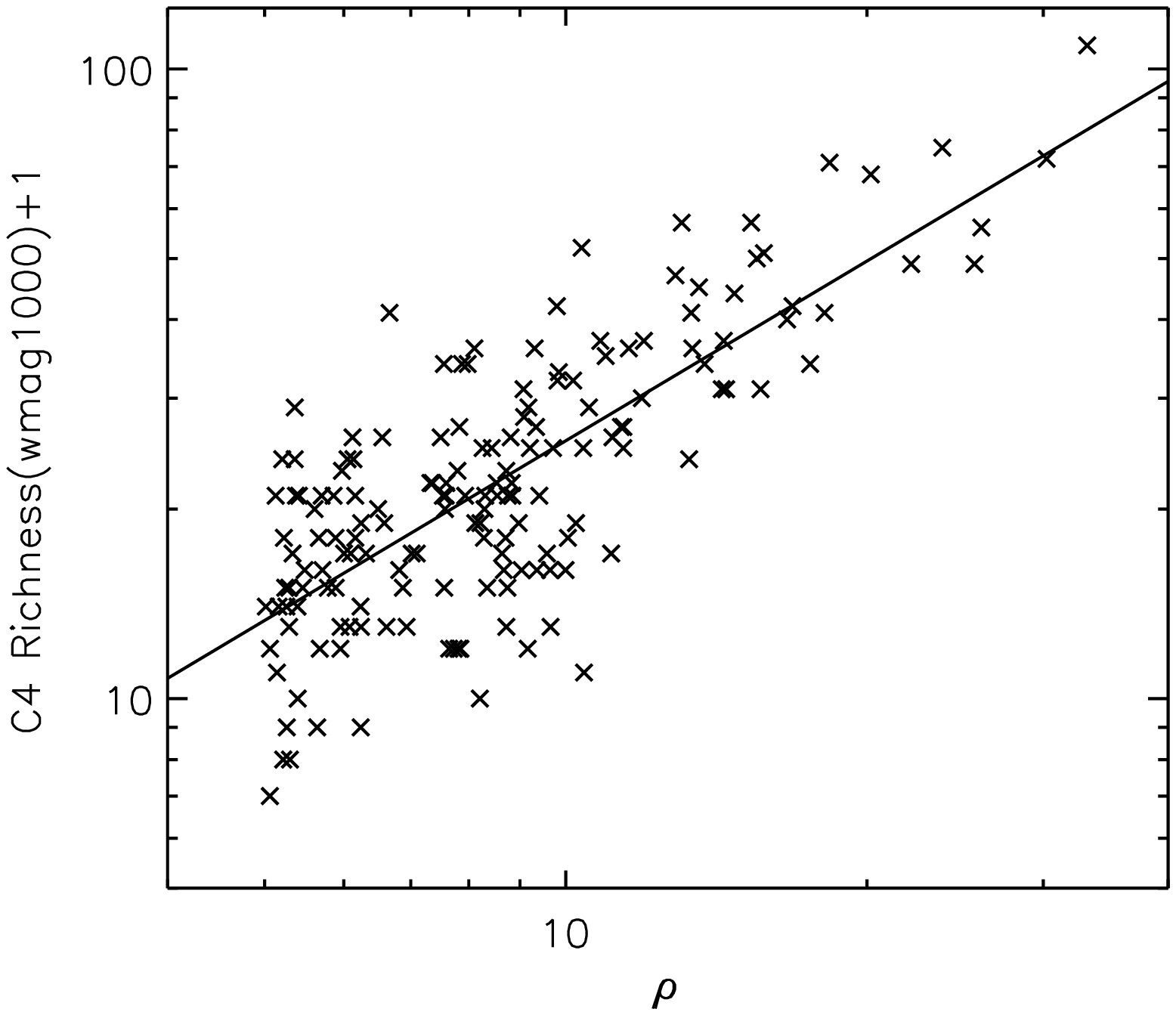}
  \caption{
Comparison between catalogues in terms of the richness of the clusters.
X-axis shows our density parameter $\rho$ (listed in Table 1)
and y-axis shows Ngal (wmag 1000) from the C4 catalogue.
The large scatter has several origins (see text for details).
The line shows the linear fit.}

 \label{C4_richness}
 \end{center}
\end{figure}

We compare our new catalogue to the C4 cluster catalogue
\citep{ab6}. While we use the density $\rho$ to represent
the richness of a cluster, the C4 Catalogue provides
the number of galaxies within various radii.
We compare our density measures within 1\,Mpc, $\rho$, with the C4 richness
(C4 parameter wmag1000) in Figure~\ref{C4_richness}.
The match is reasonably good but shows a large scatter.
The systematic difference in vertical scale has a couple of critical origins.
Most importantly, our limiting magnitude, $M_{\rm r}=-20.5$, is somewhat
brighter than that of C4 (-19.9).
Secondly, our density is not a simple count of galaxies but weighted by
the distance from the cluster center.
Thirdly, our method, unlike C4, includes additional photometric member
candidates.
Fourthly, our choice for the search radius (1\,Mpc) is smaller than
that of C4 ($1/h$\,Mpc).
Lastly, the member-search volume in C4 is a cylinder with a line-of-sight
length corresponding to four times the velocity dispersion, while ours is
an ellipsoid with a length that is 3 times the velocity dispersion.
The large scatter is mainly a result of two effects.
To begin with, the position of the central galaxy of a cluster can be
different mainly because our method includes photometric member candidates.
In addition, the redshift of a cluster can also be different.
We notice some difference in the velocity dispersion measured as well,
which also contributes to the scatter.

\subsection{Effects of photometric members}

\begin{figure}
 \begin{center}
  \vspace{5pt}
  \includegraphics[width=1.4\columnwidth]{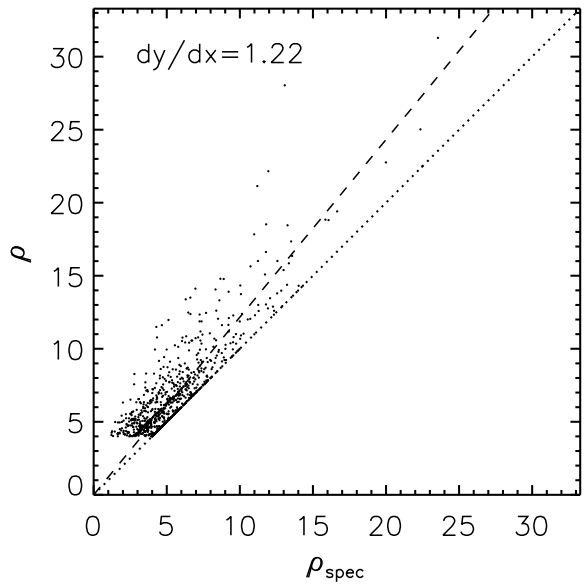}
  \caption{We compare our density measure $\rho$ with $\rho_{\rm spec}$
    (the density only with spectroscopic data).
      The dotted line presents the one-to-one relation between $\rho$ and
      $\rho_{\rm spec}$ and the dashed line shows a linear fit with the
      y-intercept fixed at the origin.
    The departure gets larger in dense regions. The density increases by $22\%$
    through the addition of photometric member galaxies.}
 \label{rho_true_all}
 \end{center}
\end{figure}

The inclusion of the CMR-selected photometric members
increases our local density measures by $\sim 22\%$
(See Figure \ref{rho_true_all}).
In this figure, the small black dots show the density measures
of our clusters based only on the SDSS database.
A linear fit to the data is shown as a dashed line
(showing the 22\% increase) and the one-to-one relation
is presented as a dotted line.

To check the validity of the CMR technique further, we attempt to recover
our spectro-photometric density $approximately$
by dividing the measured spectroscopic density $\rho_{\rm spec}$ by
the estimated spectroscopic completeness fraction $f_{\rm spec}$.
Figure~\ref{rho_true} compares $\rho_{\rm spec}/f_{\rm spec}$ (crosses)
with our spectro-photometric density $\rho$ (the upper ends of arrows).
They agree reasonably well.
Since $\rho$ is a complex parameter including
the distance information while $f_{\rm spec}$ is not, we also
make the comparison to $N_{\rm gal}$, which also shows a tight correspondence.

\subsection{Additional spectroscopy for the test}

\begin{figure}
 \begin{center}
  \vspace{5pt}
  \includegraphics[width=\columnwidth]{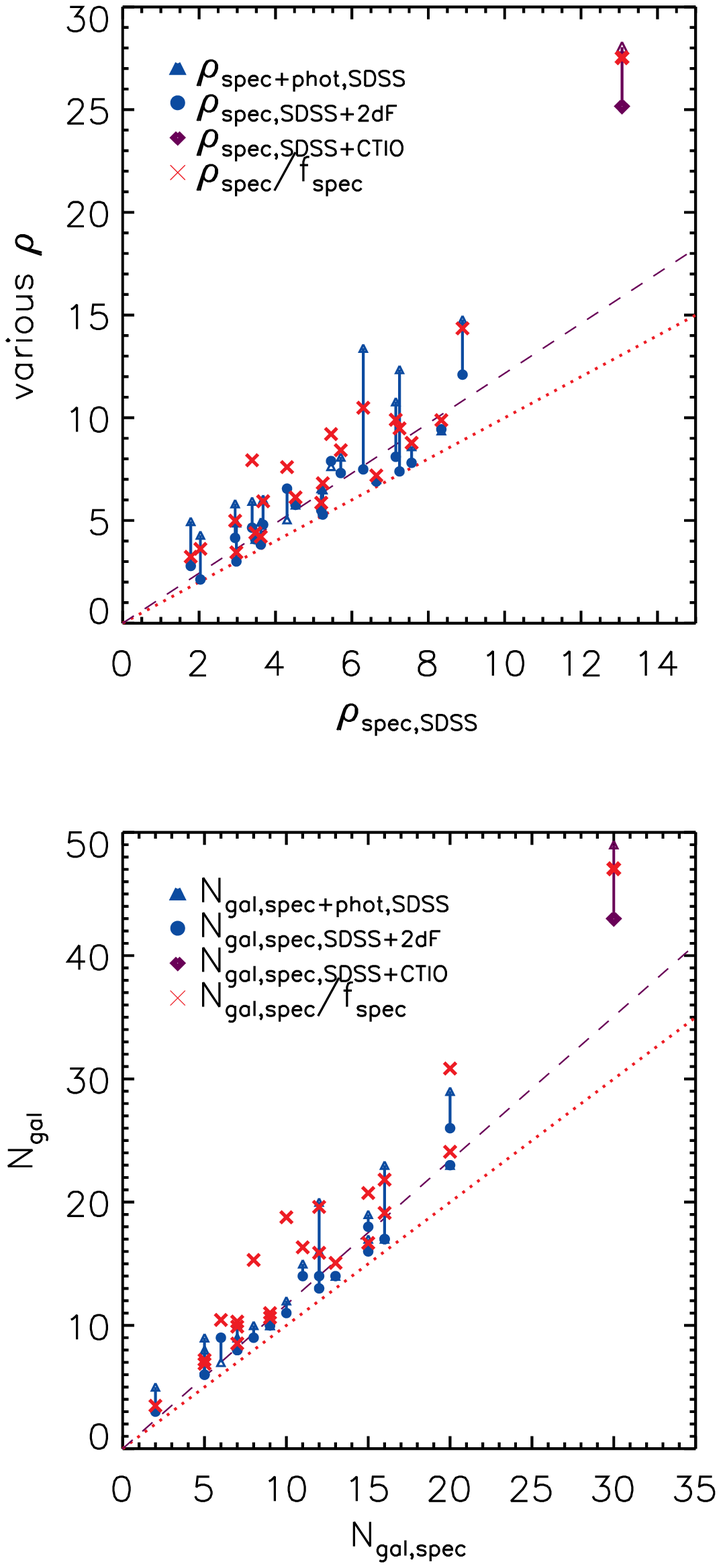}
  \caption{
$Top$: Comparison between the spectroscopy-only density measures (x axis)
and various improved density measures (y axis).
The solid triangles show our spectro-photometric densities ($\rho$).
The solid circles show the spectroscopic densities based on the SDSS
and 2dF database combined.
The purple diamond in the top right shows the spectroscopic density
of Abell\,2670 based on both the SDSS and our new CTIO database.
The crosses show the result of rough correction for the incompleteness
problem, using $\rho_{\rm spec}$ (i.e., x axis) and the completeness
fraction $f_{\rm spec}$.
$Bottom$: the same as the top panel but for the number of galaxies.
The dotted and dashed lines are the same as in Fig. 12.}
 \label{rho_true}
 \end{center}
\end{figure}

For 22 of our clusters we were able to find 2dF spectroscopic
data with which we identified further member galaxies.
Using these new spectroscopic members, we computed
new (enhanced) $\rho_{\rm spec}$ and compared them to our
SDSS-spectro-photometric $\rho$ in Figure \ref{rho_true}.
The x axis shows our 2-d SDSS-spectroscopic density ($\rho_{\rm spec}$)
that suffers from the incompleteness problem.
The y axis shows density measures improved upon $\rho_{\rm spec}$  through
various methods.
The blue filled circles show the new spectroscopic
density measures for the 22 clusters including the added
2dF member candidates: $\rho_{\rm spec,SDSS+2dF}$.
Only a few of the new SDSS+2dF spectroscopic density measures
($\rho_{\rm spec,SDSS+2dF}$) overshoot our 22\% enhancement prediction
(dashed line). Even in this case, however, the SDSS+2dF spectroscopic
density measures ($\rho_{\rm spec,SDSS+2dF}$) are still lower than our
SDSS-only spectro-photometric density measures
($\rho_{\rm spec+phot,SDSS}$; that is, our density measures in Eq. 6)
in most cases, as indicated by the arrows. The end points of the arrows
indicate our SDSS-spectro-photometric density measures.

In order to test our result, we have performed CTIO Hydra
MOS observation of Abell 2670 and significantly increased
the spectroscopic coverage of this cluster from 65\% (within 1\,Mpc)
to 92\%.  We chose this cluster as a test case because it suffers from
an extreme case of the spectroscopic incompleteness problem as
mentioned in \S 1.

The observation was performed on 1--3 December 2006 with Hydra,
a Multi-Object-Spectrograph mounted on Blanco 4m-Telescope at CTIO.
The wavelength range covers 3600--8000\AA\, and the spectral resolution is
2.3\AA/pixel with mean signal-to noise ratio of 13.
Originally all of the early-type galaxies in Abell 2670 with $r < 18$ were
selected as targets.  In order to execute our plan as completely
as possible, we used three fiber-configurations to evade fiber collision
issues in dense regions. For each fiber configuration, a total of
45 minutes which consists of 3 exposures of
15 minutes was taken for median combine and cosmic-ray removal.
All the processes are done mainly with IRAF \texttt{hydra} package.
The radial velocities of galaxies were measured by
cross-correlation method using  IRAF \texttt{rvsao} package and
cross-correlation templates of SDSS.
In the top right of Fig. \ref{rho_true}, we show our new SDSS+CTIO
spectroscopic density ($\rho_{\rm spec,SDSS+CTIO}$) at $\approx 25$
(purple diamond at the top right).
The value, based on a relatively complete (92\%) spectroscopic survey,
now is a factor of two larger than the previous $\rho_{\rm spec,SDSS}$ and
much closer to our spectro-photometric density $\rho_{\rm spec+phot,SDSS}$.

\section{Summary}

We develop and test a method for finding galaxy clusters in
the SDSS spectroscopic and photometric database. Our method
improves over the previous density measurements by finding additional cluster
member candidates via the color-magnitude relation technique. The
problem of spectroscopic incompleteness due to fiber collisions
in dense fields is minimized in our method.
The member galaxies selected by the color-magnitude relation lead to
a satisfactory completeness and purity.
With this method, we find 924 galaxy clusters of which 212 are new.
We provide a catalogue of these
clusters including important properties such as the virial radius,
velocity dispersion, and richness parameters.
The density we estimate is in good agreement with the properties
relating to the cluster mass and size.
We also provide an estimating scheme for our spectro-photometric
density measure simply using the spectroscopic completeness rate.
Our new density information on galaxy clusters could be useful for
the study of the environmental effect on the galaxy evolution.

\acknowledgements

We thank the anonymous referee for numerous suggestions and criticisms which
improved the clarity of the paper. We would like to thank
Jong-Hak Woo and Sangmo Sohn for their help in the data reduction.
This work was supported by grant No.~R01-2006-000-10716-0 from the Basic
Research Program of the Korea Science and Engineering Foundation
to the corresponding author S.K.Y.

\clearpage

\end{document}